%% file: main.tex
\documentclass[sigconf]{acmart}

\AtBeginDocument{%
  \providecommand\BibTeX{{%
    \normalfont B\kern-0.5em{\scshape i\kern-0.25em b}\kern-0.8em\TeX}}}

\usepackage{times}
\usepackage{soul}

\usepackage{url}

\usepackage{CJKutf8}

\usepackage{graphicx}
\usepackage{amsmath}
\usepackage{booktabs}
\usepackage[ruled]{algorithm2e}
\usepackage{amsfonts}
\usepackage{subcaption}
\usepackage{bmpsize}
\usepackage{graphicx}
\usepackage[export]{adjustbox}
\usepackage{dblfloatfix}
\newcommand{\rowconcat}{\text{RowConcat}}
\newcommand{\colconcat}{\text{ColumnConcat}}
\newcommand{\roll}{\text{RowCyclicPermute}}
\newcommand{\rowsplit}{\text{RowEvenSplit}}

\usepackage{latexsym} 

\title{A unified Neural Network Approach to E-Commerce Relevance Learning}
\renewcommand{\shorttitle}{A Unified DL Approach to E-Commerce Relevance}

\author{Yunjiang Jiang}
\email{yunjiang.jiang@jd.com}
\author{Yue Shang}
\email{yue.shang@jd.com}
\affiliation{%
  \institution{JD.com Silicon Valley Research Center}
  \streetaddress{675 E Middlefield Road}
  \city{Mountain View}
  \state{CA}
  \postcode{94043}
  \country{USA}
}
\author{Rui Li}
\email{richard.rui.lee@gmail.com}
\author{Wen-Yun Yang}
\email{wenyun.yang@jd.com}
\affiliation{%
  \institution{JD.com Silicon Valley Research Center}
  \streetaddress{675 E Middlefield Road}
  \city{Mountain View}
  \state{CA}
  \postcode{94043}
  \country{USA}
}
\author{Guoyu Tang}
\email{tangguoyu@jd.com}
\author{Chaoyi Ma}
\email{machaoyi@jd.com}
\affiliation{%
  \institution{JD.com}
  \streetaddress{JD Building, No. 18 Kechuang 11 Street, BDA}
  \city{Beijing}
  \postcode{101111}
  \country{People's Republic of China}
}
\author{Yun Xiao}
\email{xiaoyun1@jd.com}
\author{Eric Zhao}
\email{ericzhao@jd.com}
\affiliation{%
  \institution{JD.com Silicon Valley Research Center}
  \streetaddress{675 E Middlefield Road}
  \city{Mountain View}
  \state{CA}
  \postcode{94043}
  \country{USA}
}
\renewcommand{\shortauthors}{Jiang and Shang, et al.}

\copyrightyear{2019} 
\acmYear{2019} 
\setcopyright{acmcopyright}
\begin{document}

\acmConference[DLP-KDD'19]{1st International Workshop on Deep Learning Practice for High-Dimensional Sparse Data}{August 5, 2019}{Anchorage, AK, USA} 

\acmPrice{15.00}
\acmDOI{10.1145/3326937.3341259}
\acmISBN{978-1-4503-6783-7/19/08}

\begin{abstract}
Result relevance scoring is critical to e-commerce search user experience. Traditional information retrieval methods focus on keyword matching and hand-crafted or counting-based numeric features, with limited understanding of item semantic relevance. We describe a highly-scalable feed-forward neural model to provide relevance score for (query, item) pairs, using only user query and item title as features, and both user click feedback as well as limited human ratings as labels. Several general enhancements were applied to further optimize eval/test metrics, including Siamese pairwise architecture, random batch negative co-training, and point-wise fine-tuning. We found significant improvement over GBDT baseline as well as several off-the-shelf deep-learning baselines on an independently constructed ratings dataset. The GBDT model relies on 10 times more features. We also present metrics for select subset combinations of techniques mentioned above.     
\end{abstract}
\maketitle

\input{introduction}

\input{related}

\input{model}
\input{experiments}

\input{case_studies}
\section{Conclusion}
We present a complete description of a 2-stage neural net based relevance scoring system. On an independent test set our model, using only a single text feature (item title), outperforms traditional IR systems based on 60 numerical features by a wide margin, and allows generalization into the long tail where only text features are available. By exploiting several generic architectural enhancements, most notably batch negative co-training, we obtain significant qualitative improvement over the base network. Though we only study feed-forward architectures here for the sake of serving scalability, these techniques can also apply to more computationally intensive ones such as CNN/RNN/attention.

% \newpage
%% \small
\bibliographystyle{abbrv}
\bibliography{main}

\end{document}

%% file: introduction.tex
\section{Introduction}
%Online shopping has become ubiquitous in the modern world. With the rise of US platforms like Amazon, Ebay, as well as their Chinese counterparts JD and Taobao, many traditional shopping tasks have been streamlined to a series of swipes and clicks. Instead of a shopping catalogue, or an hour-long stroll through a physical store, customers are now presented a search box in a web page to quickly zero in on the items they intend to buy.
%This technological innovation puts an undue burden on the quality of the e-commerce search engine, who needs to provide results for myriad search queries, many of which are ill-formed, incomplete/ambiguous, or simply not corresponding to a real product. Even more serious users enter search queries with words that do not necessarily conform to the standard retail jargons. Thus keyword/key-phrase matching may not capture the meaning accurately, or return any result at all. This is where embedding based deep learning models can come in handy. 

Online shopping has become ubiquitous in the modern world. A good e-commerce search engine, which helps users quickly zero in on their intended products from billions of candidates, is thus essential to users' online shopping experiences.

The most fundamental problem in e-commerce search is learning relevance between query and items. Compared with traditional web search or document retrieval, relevance learning in e-commerce search has two unique requirements. 
\begin{itemize}
    \item Unlike web documents, e-commerce items typically have little business-side textual content, mainly a short title of less than 50 characters; user comments are less reliable and typically more skewed towards older products. Thus simple keyword-based matching may not capture semantic matches effectively. We observe that even serious users enter search queries with words that do not necessarily conform to standard retail jargons. Thus, an ideal model should be able to do fuzzy match between queries and product titles.   
    \item Due to the profit-making nature of e-commerce, personalized recommendation and other promotional logic often play a more important role than relevance in final ranking. On the other hand, irrelevant items should clearly be filtered. An ideal relevance model therefore should calibrate well with an established relevance grading, such as a binary classification for relevant and irrelevant ones or PEGFB scale ratings (Perfect, Excellent Good, Fair, Bad).  
    % Relevance scoring is thus relegated to a filtering role. Second, in e-commerce search, results are ranked according to predicted conversation/click-through rate with consideration of relevance score, and the relevance score is mainly used for filtering irrelevant candidates in ranking as well as in other business logic. 
\end{itemize}

In this paper, we share our experiences of learning such an ideal relevance model for e-commerce search. We choose a deep neural network based approach as it can learn latent semantics based on raw text features. As a deep model usually requires a huge amount of data to train and human labels are expensive, we explore two sources of label supervision: 1) \textbf{user clicks}, an indirect signal for relevance (i.e., users clicks are affected by many factors such as relevance, price, and sale volumes), and 2) limited \textbf{editorial relevance ratings}. Our main contribution is to demonstrate three highly general techniques to enhance the quality of such a hybrid deep neural network, to achieve accurate relevance prediction.

\begin{figure*}
\begin{subfigure}[b]{0.45\textwidth}
\includegraphics[width=1.3\textwidth, left]{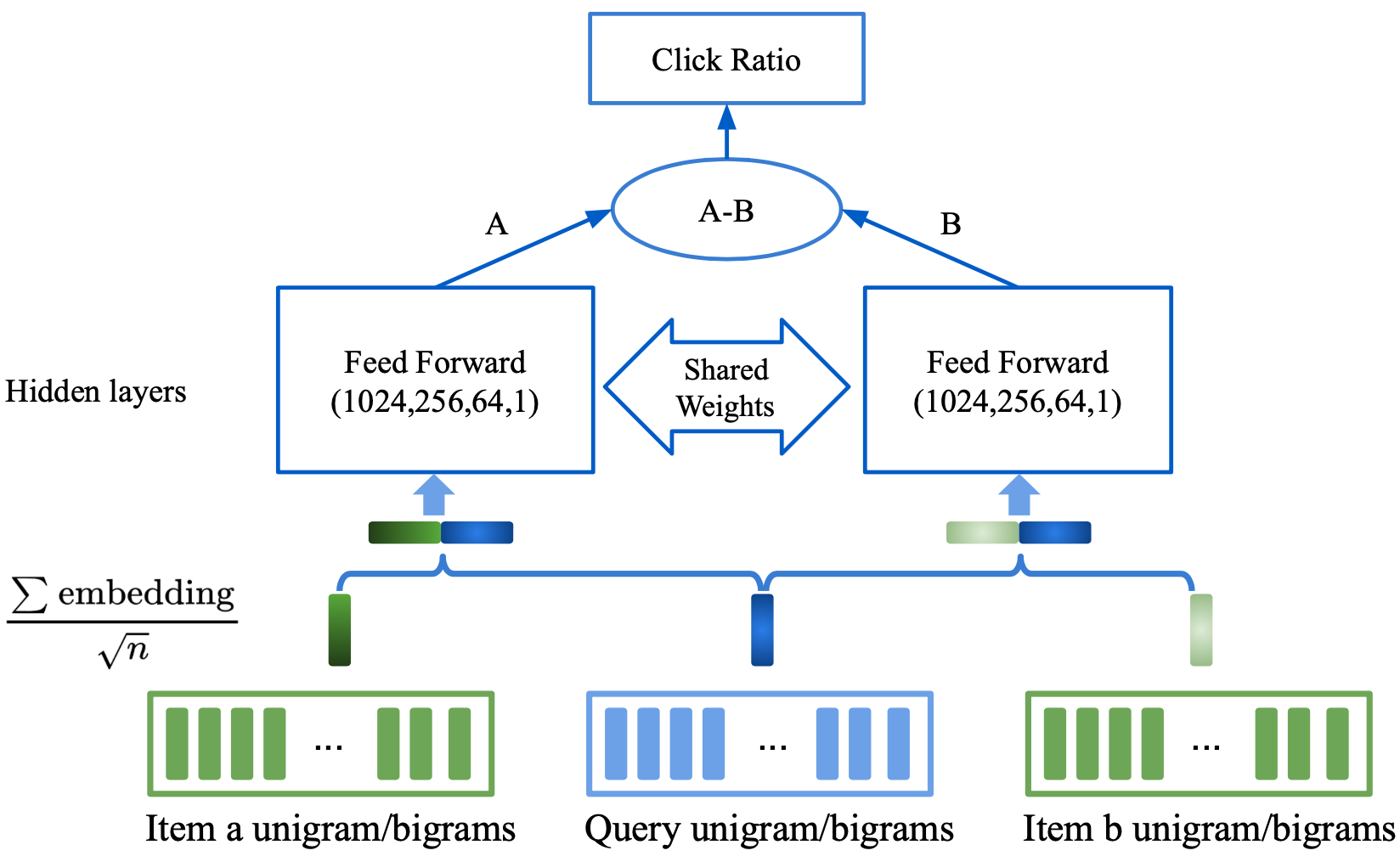}
\caption{Pairwise Siamese}
\label{fig:Siamese-network}
\end{subfigure}\hfill
\begin{subfigure}[b]{0.45\textwidth}
\includegraphics[width=0.85\textwidth, right]{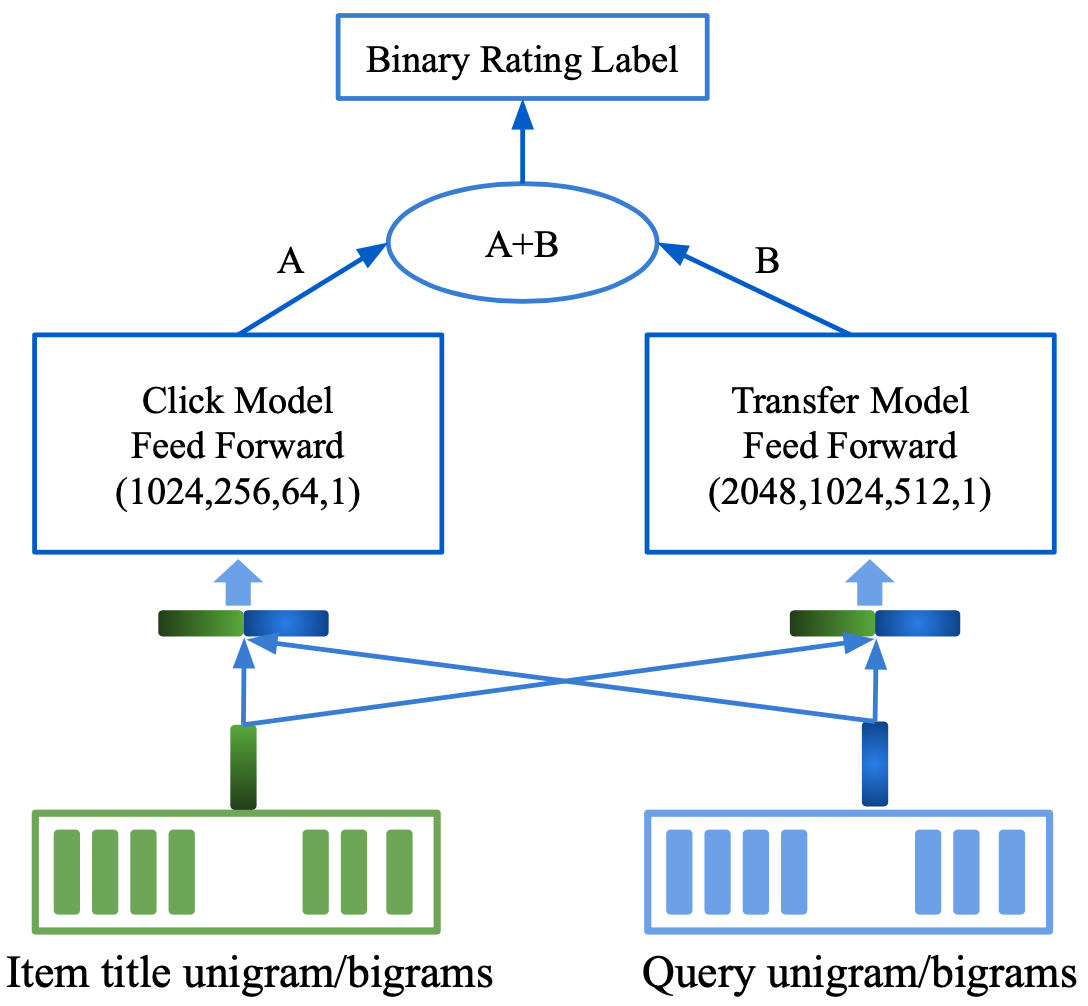}
\caption{Point-wise ensemble fine-tuning}
\label{fig:transfer-network}
\end{subfigure}\hfill
\caption{Model Architectures}
\label{fig:networks}
\end{figure*}

%The task of search relevance scoring has a close relationship with natural language processing. While text features such as result description, result title, etc, are clearly very important for a human to judge relevance, other features such as aggregated user feedback, page rank, etc, also play a fundamental role in determining the overall relevance score. Our internal system uses an intricate combination of both types of features.

%Our main contribution is to demonstrate the use of several highly general techniques to enhance the quality of the most basic feed-forward neural network, supervised by a combination of a large amount of noisy user clicks and limited editorial labels.

\subsection{Siamese Network for Pairwise Relevance}
While user clicks are significantly correlated with result relevance \cite{huang2013learning,neural_ir_book}, click derived signals, such as CTR, are unsuitable as a point-wise learning target for relevance, especially in e-commerce, because 1) they are not well-calibrated to absolute relevance ratings, such as the widely used PEGFB scale, and 2) point-wise click labels have very skewed distribution (typically concentrated at 0), making it easy for the model to ``cheat". However, clicks in aggregate are invaluable at distinguishing between good and bad results under the same query. For this reason, we choose a pairwise architecture that learns preference between a pair of items under the same query. To enable scoring single items during serving, the model takes a Siamese form \cite{bromley1994signature}: the final logit for the item pair is expressed as the difference of two individual item scores from two towers of shared weights, and only one of the two identical towers is used during serving. While most applications of Siamese network in the literature are for pairwise similarity learning \cite{chopra2005learning,yin2016abcnn}, we use it for pairwise discrimination. Experiments show that this 2-tower setup actually beats the non-Siamese pairwise baseline on several key metrics (Table~\ref{tbl:exp1}).

\subsection{Efficient Training with Batch Negatives}
% those returned to users or users have seen, which are usually somewhat relevant to given queries. Thus, our model is trained to distinguish between ``relevant'' and  ``less relevant'', but not ``relevant'' or ``irrelevant''. E.g., we observe that a vanilla Siamese model give a high score for an item, which has a high CTR for a query (e.g., cellphone) high for another random query (e.g., paper cup), as we does not give any supervision for the model to punish this random item.

One challenge we face in using clicks as training labels is the paucity of negative relevance examples. To ensure good user experience, top items in a query session are usually at least somewhat relevant. Thus the model rarely sees completely unrelated (query, item) pairs. This causes a serious distributional skew during serving, where we intend to score thousands of relevant and irrelevant items for a query. As a remedy, we design an efficient batch negative algorithm (Algo~\ref{batch-negative-forward-pass}), to supplement the original $n$ session pairs in each batch with $n(n-1)$ additional random negative (query, item) pairs. Intuitively, this gives the model a more global view of possible items for each query. Amazingly all key metrics on the original examples improve (Table~\ref{tbl:exp1}). Case study also shows that unrelated items now score much lower under the batch negative enhancement.

%A related challenge is the prevalence of click-baits or universally popular items in the search log. Due to the relative simplicity and focus of e-commerce queries compared to say web search, and the way traditional keyword based retrieval works, results tend to form disjoint clusters around semantically similar queries, making it difficult for the model to generalize across queries. As a consequence, the vanilla Siamese model above often behaves in a query-independent manner, producing very similar scores for wildly differing queries. To remedy this undesired property, we introduce an efficient batch negative algorithm, that supplements the original $n$ session pairs in each batch with $n(n-1)$ additional random negative pairs. Somewhat surprisingly, this gives additional metric boost to the Siamese model. More importantly, case study shows that unrelated items now score much lower under the batch negative enhancement.

%In addition, the conjunction of session and batch negatives mitigates the data skew between training and serving. In production serving, we intend to score thousands of items per query using the neural net model. During training, however, we only collect items that the user sees, which typi. Thus the model may never see some items during training. Including batch negatives examples gives the model a more global view of possible items for each query.

\subsection{Fine-Tuning with Human Supervision}
% Thus, we apply transfer learning to the click model, with labels in the PEGFB scale provided by human editors, to introduce limited human supervision.

Lastly, we fine-tune the model with editorial labels, in a \textbf{point-wise} fashion. This is needed for two reasons:
\begin{itemize}
    \item clicks, though abundant, can only approximate result relevance, and are often misleading: a user can click on a result simply because of an attractive title or being part of a malicious click farm to manipulate search ranking.
    \item the ultimate system goal is to filter irrelevant items (those below the \textbf{Good} rating), but \textbf{pairwise} models are not well calibrated to an absolute notion of relevance; e.g., a score of 0.0 could mean Excellent or Fair depending on query, because the Siamese model is shift invariant.
\end{itemize}
Point-wise architecture at this step is conveniently adapted from a single tower of the base Siamese model. This leads to an additional 6\% gain in eval AUC.

%% file: related.tex
\section{Related Works}

In this section, we review related work on relevance learning, which aims to learn a function to determine relevance between a query and document pair. We categorize these models as traditional learning to rank based methods and deep learning based methods.

\subsection{Traditional Learning to Rank Methods}
Due to the importance of search engines, using model machine learning techniques to improve relevance ranking, often referred as learning to rank, has been an active research topic in the past few decades. A lot of progresses have been made, including point-wise models like RankNet~\cite{L2RPoint}, pairwise models like RankSVM~\cite{ranksvm} and GBRank~\cite{gbrank}), and list-wise models like AdaRank~\cite{AdaRank} and LambdaMart~\cite{LambdaMart}); see \cite{l2rSurvey} for a complete survey. Here we discuss some key results that our work builds upon. First is using pairwise relevance preference instead of absolute relevance grade as the learning target introduced in~\cite{ranksvm,gbrank}; this is the basis of the Siamese pairwise architecture. We do not use a list-wise approach for optimizing ranking results globally, because unlike web search, relevance is only a signal in final ranking and is mainly used as filtering in e-commerce search. Second is exploiting easy-to-get but noisy click logs instead of expensive but accurate human labels to learn relevance \cite{ranksvm}. Our model actually leverages both labels. All the traditional methods rely on manually defined features, such as BM25, matching positions and page-rank, and cannot effectively utilize raw text features. Thus, they often fail to evaluate \emph{semantic} relevance if a document does not contain exact terms in a query.

\subsection{Neural Network Based Ranking Methods}
In the past few years, a large number of embedding-based neural network models have been successfully applied to learn semantic relevance between queries and documents; \cite{INR-061} provides an excellent survey. These works are roughly divided into two camps: using neural network induced embeddings as features in a 2-stage process, or directly in an end-to-end manner.

In the first category, \cite{van2016learning} learns embeddings as features for a final ranking objective. \cite{yu2014latent} uses the embeddings to promote result diversity. The resulting embeddings are amenable to fast online retrieval and similarity analysis. \cite{ai2017learning} in addition learns user embeddings for personalized search.

In the second category, DSSM \cite{huang2013learning} and their followup work CDSSM \cite{cdssm} pioneered the application of deep neural networks in end-to-end relevance learning. Further, new models, such as DRMM \cite{guo2016deep}, Duet \cite{duet}, DeepRank \cite{deep_rank} have been proposed to improve content based models via exploring traditional IR lexical matching signals (e.g., query terms importance, exact matching) in neural networks. This is the approach that we follow. Also we keep the user dimension out of the equation to stay focused on pure relevance learning.

DSSM uses two separate feed-forward towers, also referred to as Siamese network, to encode query and result as embeddings, and optimizes a softmax loss based on the cosine distance of those embeddings. Our notion of Siamese network is different. Each of our training examples consists of a query and a pair of items, where the model allows interaction between the query and either item, but no interaction between the two items (see Figure \ref{fig:networks}\subref{fig:Siamese-network}). We note that, our model can be easily extended to include those new signals beyond just query and title, or new architectures (e.g., RNN, Transformer) in each tower of the Siamese network. Our main contributions are three general techniques: 1) Siamese network for learning pairwise preference, 2) efficient training with batch negatives, and 3) point-wise fine-tuning via human labels.

%% file: model.tex
\section{Model Description}
% \label{sec:model}
Much of the relevance ranking literature takes a single query and multiple documents as input, such as DSSM and its variants. This is well suited for the ranking problem. Our ultimate use case, however, is to filter out irrelevant documents, based on a score and a global threshold. In other words, we learn a relevance classification instead of a ranking model.

For simplicity and portability, we use title text as the only item side feature. In practice, numeric features can be sparse, especially for tail queries, whereas item titles are almost always available, and can help better generalize from torso queries. Our model is first trained on user clicks, then fine-tuned using human labels.

\subsection{Siamese Pairwise Network}

Due to the abundance of clicks in search log, compared to other user signals such as purchase or placement into shopping cart, we choose clicks as weak labels in the initial model warm-up step.

Since clicks are most informative of relative relevance, we first learn preference between a pair of items under each query: each training example consists of the triple ($Q$, $\rm{item}_a$, $\rm{item}_b$), where $Q$ stands for query, and $\rm{item}_a$ and $\rm{item}_b$ are two co-occurring items under $Q$, swapped randomly. We call such triples \textbf{session pairs}.

The goal of the model is to predict the click ratio of $\rm{item}_a$, as a proxy for its relative relevance, that is, we train a classifier $F$ according to

\begin{equation}
    F(Q, \rm{item}_a, \rm{item}_b) \sim \frac{\rm{click\_count}_a}{(\rm{click\_count}_a + \rm{click\_count}_b)}
\end{equation}

The obvious simpler alternative is to instead fit a function G to the pointwise click-through rate objective
\begin{equation}
    G(Q, \rm{item}) \sim \frac{\rm{click\_count}}{\rm{view\_count}}
\end{equation}

However this approach suffers badly from label imbalance in the e-commerce search setting, where the click through labels are typically very sparse and have a highly skewed asymmetric distribution. For instance, letting $G$ be identically 0 would already achieve a reasonably good logloss or square loss, which is completely useless for relevance scoring.

By randomly swapping $\rm{item}_a$ and $\rm{item}_b$, the function F cannot rely on any distributional prior to beat random guessing. Since the problem can be viewed as binary classification, logloss is a natural choice.

For the convenience of scoring a single item during serving, we further restrict F to the factorized Siamese form:
\begin{align}
F(Q, \rm{item}_a, \rm{item}_b) = H(Q, \rm{item}_a) - H(Q, \rm{item}_b)
\end{align}
Taking difference seems to work reasonable well compared to other more elaborate contrastive functions. Since the click ratio labels are not necessarily in $\{0, 1\}$, hinge loss does not apply meaningfully to the within-session pair examples. However it does make sense to apply hinge loss to the batch negative sampled pairs. We found the hybrid session logistic-loss and batch negative hinge-loss approach to improve eval logistic-loss and auc significantly on the session pairs. However it did not improve key metrics during the human-labelled fine-tuning step.

The single tower function H computes the following:
\begin{enumerate}
    \item For both query and item, we sum up the unigram/bigram embedding vectors (of the same dimension), 
    divided by sqrt(n), then concatenated to form a 2d tensor E of shape (n, 2d), 
    where n is the training/eval batch size, and d is the embedding dimension. 
    \item The prediction in logit space is then calculated as $\rm{H}'(E)$, where $\rm{H}'$ is a feed-forward neural net, 
    consisting of 3 ReLU layers [1024, 256, 64, 1].
\end{enumerate}

\subsection{Efficient Training with Batch Negatives}

\begin{algorithm}[!b]
\DontPrintSemicolon
\SetKw{KwWhere}{where}
\SetKwInOut{Input}{input}\SetKwInOut{Output}{output}
\Input{Batch size: $n$}
\Input{Feed-Forward network: $H'$}
\Input{Batched embedding tensors for queries, positive items, negative items: $Q$, $I_+$, $I_-$}
\Output{Logits for the original and batch negative pairs}
$E_+ \leftarrow \colconcat(Q, I_+)$\;
$E_- \leftarrow \colconcat(Q, I_-)$\;
$E \leftarrow \rowconcat(E_+, E_-)$\;
\ForEach{$1 <= i <= n$} {
    $I_+ \leftarrow \roll(I_+)$\;
    $E_{\rm{BN}} \leftarrow \rowconcat(Q, I_+)$\;
    $E \leftarrow \rowconcat(E, E_{\rm{BN}})$
}
$\vec{P} \leftarrow \rowsplit(H'(E), n+1)$\;
original logits $\leftarrow P_1 - P_2$\;
BN logits $\leftarrow \rowconcat(P_3 - P_1, \ldots, P_{n+1} - P_1)$\;
All logits $\leftarrow \rowconcat(\text{original logits}, \text{BN logits})$

\Return{$\rm{All}\text{ }\rm{logits}$}
\caption{Batch negative augmented forward pass.}
\label{batch-negative-forward-pass}
\end{algorithm}

The Siamese pairwise model trained using session pairs only, as described in the previous section, already exceeds DSSM on several evaluation metrics (Table~\ref{tbl:exp1}). More careful case study, however, reveals a fundamental problem. Many items have little variation in their scores regardless of which query is used. Clearly query is an important input to any relevance model, and our goal is not to create a query-independent item quality model.

One explanation is that unlike web search, e-commerce item attractiveness often plays a much more important role than relevance in determining users’ feedback. Truly irrelevant items (negative examples) are also rare in a session, since our search system has optimized its relevance for years and most of returned items are already relevant.

To address these two problems, we borrow ideas from word2vec \cite{word2vec}, and augment the training data with $O(n^2)$ soft negative pairs, constructed within the same batch. In the simplest point-wise setting, we pair up the query of each example in the batch with items of other examples. In the pairwise case, a natural modification is to pair the query and positive item of each example with the positive items of other examples. The motivation is that since the training set is globally uniformly shuffled, the chance that two examples within a single mini-batch have the same query, or even synonymous queries, is extremely low. Thus we can assign with high confidence negative labels to each new pair. 

For $a, b$ two items co-occurring under a query $q$, let $\rm{click}(a; b, q)$ stand for the number of clicks item $a$ received in the context of $q$ and $b$. 
Further, denote by $\Lambda$ the set of all such co-occurring triples $(a, b, q)$ within a mini-batch, ordered by $\rm{click}(a; b, q) \geq \rm{click}(b; a, q)$, followed by alphabetic tie-breaking of the item titles.

Then the mini-batch training objective takes the form
$\rm{train loss} =\text{original loss} + \text{BN loss}$, where 

\begin{align}
\text{original loss} &= \sum_{(a, b, q) \in \Lambda} \lambda(a, b, q; \ell(a, b, q)) \\ % logloss(logit(a, q) - logit(b, q), \textit{label}) 
\text{BN loss} &= \sum_{(a, b, q), (a', b', q') \in \Lambda, q' \neq q} \lambda(a, a', q; 0) \\ % logloss(logit(a, q’) - logit(b, q’), 0) 
\ell(a, b, q) &= \frac{\rm{click}(a; b, q)}{\rm{click}(a; b, q) + \rm{click}(b; a, q)} \\
\lambda(a, b, q; \ell) &= \tau(\rm{logit}(a, q) - \rm{logit}(b, q), \ell),
\end{align}
and
\begin{align}
\tau(x, \ell) = -\ell \log \sigma(x) - (1 - \ell) \log \sigma(-x)
\end{align}
is the logloss function for logit $x \in \mathbb{R}$ and label $\ell \in [0, 1]$. Here $\sigma(x) := \frac{1}{1 + e^{-x}}$ is the sigmoid function.

Thus, the batch negative (BN) loss is the sum of $n (n - 1)$ terms, where $n$ is the batch size, 
because each positive item can be paired with all positive items from the other $n - 1$ examples. 
The relative weights between the two loss components do not seem to matter, as judged by their final eval metrics. Algorithm~\ref{batch-negative-forward-pass} show the implementation of computing the forward pass with $O(n^2)$ batch negative examples efficiently within the tensorflow framework.

Geometrically, with batch negatives, the query embeddings are not allowed to cluster around a single direction, since unrelated (query, item) pairs must score much lower than related pairs.

On the 6 month data, batch negative AUC reaches 0.99 within 100k steps of 128 batch size; in contrast, original example AUC takes 6 million steps to converge. That is because the batch is now dominated by $O(n^2)$ soft negative examples, which are easy to learn. The inclusion of batch negative examples has little effect on how quickly metrics converge on the original examples, making weight tuning unnecessary.

% \begin{figure*}
% \begin{subfigure}[b]{0.45\textwidth}
% \includegraphics[width=1.2\textwidth, left]{SiameseDiagram.png}
% \caption{Pairwise Siamese}
% \label{fig:Siamese-network}
% \end{subfigure}\hfill
% \begin{subfigure}[b]{0.45\textwidth}
% \includegraphics[width=0.85\textwidth, right]{PointwiseTransferLearning.eps}
% \caption{Pointwise transfer learning}
% \label{fig:transfer-network}
% \end{subfigure}\hfill
% \caption{Model Architectures}
% \label{fig:networks}
% \end{figure*}

\subsection{Fine Tuning via Relevance Ratings Data}

% While click feedback is abundant for the scale required by neural network, it is only a proxy for result relevance. Often a user clicks on a result simply because of a universally attractive title. In other situations malicious users hire click farm to manipulate search ranking. E-commerce users also tend to click on cheap items, regardless of its exact utility. Thus it is natural to fine-tune the model using human relevance ratings.

In the last stage, we use human labeled data to fine-tune the click model. Since rating millions of items is prohibitively expensive, it is impractical to train a model from scratch. Instead we take the click model output and its word embeddings as a starting point, and learn a new set of hidden layers from random initialization. In order to assign absolute relevance grade to each (query, item) pair, we also switch from pairwise to a point-wise architecture, with binary labels determined by whether the rating exceeds a certain threshold: Perfect/Excellent/Good ratings are considered positive; see Figure \ref{fig:networks} (\subref{fig:transfer-network}) for the model architecture. Finally we add the new point-wise network to the original click point-wise prediction, as a form of ensembling. 

The point-wise transfer learned model thus predicts the probability whether a (query, result) pair has a relevance score above Fair (2). More generally, experiments (Table~\ref{tbl:exp2}) show that the point-wise approach is better suited for optimizing Precision/Recall AUC for both positive and negative examples.

%% file: experiments.tex
% % !TEX encoding = UTF-8 Unicode

\begin{table*}[!t]
\begin{tabular}{llllllll}
\toprule
models \textbackslash\, metrics & Eval AUC & Pair Accu & Neg PR AUC & PR AUC & NDCG@10 & Mean Avg Prec & Prec@3   \\
\hline
Click-Trained Model (part)  & \textbf{0.5846}    & \textbf{0.6645}    & \textbf{0.6065}     & \textbf{0.8316}    & \textbf{0.8947}   & \textbf{0.8721} & \textbf{0.7351}     \\ 
w/o Negative Sampling     & 0.5825    & 0.6476    & 0.5968     & 0.8314    & 0.8912   & 0.8680 & 0.7299 \\
non-Siamese Pairwise & 0.5751    & 0.6071   &  - & -   &  - &  - &  -  \\  
DSSM            & 0.5692    & 0.5862    & 0.5288     & 0.6495    & 0.8681   & 0.8615 & 0.7114      \\ 
\bottomrule
\end{tabular}
\caption{Click model comparison with simplified versions}
\label{tbl:exp1}
\end{table*}

\begin{table*}[!t]
\begin{tabular}{lllllllll}
\toprule
models \textbackslash\, metrics  
& \begin{tabular}{@{}c@{}}Eval AUC\end{tabular}
& \begin{tabular}{@{}c@{}}Pair Accu\end{tabular}
& \begin{tabular}{@{}c@{}}Neg PR AUC\end{tabular}
& \begin{tabular}{@{}c@{}}PR AUC  \end{tabular}
& NDCG@10 & Mean Avg Prec & Prec@3    \\
\hline
GBDT &    -  & 0.6778     & 0.6463     & 0.7759  & 0.8995  & 0.8933 & 0.7225 \\ 
% \hline
\text{Click-Trained Model (full)} & 0.7572      & 0.6822    & 0.4832  & 0.7553 & 0.9053  & 0.9054 & 0.7240 \\
% \hline
 \begin{tabular}[c]{@{}l@{}}+Pointwise simple \end{tabular}  
 & 0.8201   & 0.6678  & 0.6598  & 0.7656  & 0.8956  & 0.8942 & 0.7220 \\ 
 % \hline
\begin{tabular}[c]{@{}l@{}}+Pairwise ensemble\end{tabular}  
& 0.7744   & \textbf{0.6956}  & 0.6546   & 0.7673  & \textbf{0.9077}  & \textbf{0.9094} & 0.7232 \\ 
% \hline
\begin{tabular}[c]{@{}l@{}}+Pointwise ensemble\end{tabular} 
& \textbf{0.8262}      & 0.6921    & \textbf{0.6698}   & \textbf{0.7779}     & 0.9023  & 0.9003 & \textbf{0.7244} \\ 
\bottomrule
\end{tabular}
\caption{Model compare results with GBDT and difference transfer learning configurations}
\label{tbl:exp2}
\end{table*}

% Case study table move here for layout
\begin{table*}[!ht]
\begin{CJK*}{UTF8}{gbsn}
\centering
\begin{tabular}{llllll}
\toprule
\multicolumn{1}{c}{} & \multicolumn{2}{c}{No Batch Negative Model} & \multicolumn{2}{c}{With Batch Negative Model}\\
\cline{2-5}
Item title  & ``cellphone''  & ``paper cup'' & ``cellphone''  & ``paper cup''   \\
\hline
\begin{tabular}[c]{@{}l@{}} 荣耀10 64GB 渐变蓝 双摄像头 双卡双待 4G全面屏手机\\
(Honor 10 64GB Blue Dual Camera Dual Sim Full Display) \end{tabular}   & 1.6292  & 1.2842  &  4.6584 & -8,2954 \\ \hline
\begin{tabular}[c]{@{}l@{}} Apple iPhone X 64GB 深空灰色 \\
(Apple iPhone X 64GB Space Grey) \end{tabular}  &   1.5081    & 1.1767 & 4.8843    & -7.1850 &\\ \hline
\begin{tabular}[c]{@{}l@{}} 荣耀9i 4GB+64GB 幻夜黑 4G全面屏手机 双卡双待\\
(Honor 9i 4GB+64GB Black 4G Full Display Dual SIM) \end{tabular} &  1.5014   &1.0814 &  4.7038   & -9.0472 & \\
\bottomrule
\end{tabular}
\caption{Siamese pairwise click model \textbf{with}/\textbf{without} batch negative for the queries ``cellphone'' and ``paper cup''}
\label{case-studies}
\end{CJK*}
\end{table*}
% case study table

\section{Experiments}

We present two sets of experiments. The first set compares models trained on 10\% of 6 month session log (about 50m examples), and 1 million unigram/bigram vocab. No human labels are used here. The second set focuses on the effect of human label transfer learning. The common base model here is trained on the full 6 month of user click data, with a vocab  size of about 16 million.
\label{sec:experiments}
\subsection{Experiments Setup}

\subsubsection{Click Data Generation}
Our primary training data consists of user click-through on search session items. We perform several compression steps to strike a balance between session coverage and data preservation:
\begin{enumerate}
	\item From each search session, generate (query, $\rm{item}_a$, $\rm{item}_b$, $\rm{is\_clicked}_a$, $\rm{is\_clicked}_b$) 5-tuples, where at least one item is clicked and below the other item in displayed position.
	\item Over a period of 180 days, aggregate the click counts to obtain the 5-tuple's (query, $\rm{item}_a$, $\rm{item}_b$, $\rm{click\_cnt}_a$, $\rm{click\_cnt}_b$).
	\item For each query, we retain at most top 100 5-tuple's of the previous step, sorted by $\rm{click\_cnt}_a + \rm{click\_cnt}_b$.
\end{enumerate}
The click data is randomly shuffled by query, and we use 90\% for training and 10\% for eval.

\subsubsection{Rating Data Generation}
We collect human labelled data totaling about 300k (query, item) pairs. On average, each query has about 5 items. The queries are shuffled at random and split into 65\% training, 30\% eval and 5\% test, from which we construct the pointwise datasets. The training and eval data sets are used for transfer learning. All the experiments are evaluated on test set. 

% \begin{table}[h]
%     \centering
%     \begin{tabular}{lll}
%     \toprule
%          Name & \#records & \#distinct queries \\ \hline
%          train set & 200k & 15k \\ 
%          eval set & 90k & 7k \\
%          test set & 13k & 3.4k  \\
%     \bottomrule
%     \end{tabular}
%     \caption{Data Set Statistics}
%     \label{tab:dataset}
% \end{table}

% \subsection{Baselines Methods} 

% {\flushleft\textbf{Gradient Boosted Decision Tree}}(GBDT) is one of the most popular non-DL based models used for learning relevance in search systems; see \cite{yin2016ranking} for detail. This is our in-house ranking model baseline, with 60 features covering a variety of user feedback signals, hand-picked matching features, including BM25, window size, etc.

% {\flushleft\textbf{DSSM}} is a state-of-the-art deep matching model for web search. We implement a query/item dot-product DSSM model, similar to the one described in \cite{huang2013learning}. 

% {\flushleft\textbf{Pairwise fully connected model}} allows full interaction among the query and the two items during training, but can only compare two items at serving.

\subsection{Evaluation Metrics}
Besides the standard IR metrics NDCG@10, MAP, and Precision@3, we calculate pairwise accuracy (Pair Accu) for two items under the same query and Precision/recall AUC (PR-AUC) for both positive and negative labels. PR-AUC is more discriminating than ROC-AUC for tasks with a large skew in the class distribution \cite{davis2006relationship}. In addition, it's a natural generalization of classification accuracy, which we care about for the filtering task. All experiments are evaluated on a 13k (query, item) pair test set. 

% \textbf{Pairwise Accuracy}:For each query session, we list all the item pair combinations, and evaluate how well the model scores preserve label orderings of pairs of items under the same queries. 
% \textbf{precision-recall AUC}: precision-recall AUC(PR-AUC) is the area under precision-recall curve. PR-curves have been cited as an alternative to ROC curves for tasks with a large skew in the class distribution\cite{davis2006relationship}. PR-AUC will calculate a comprehensive score of precision and recall.

\subsection{Experiment Results}
According to the previous discussion, we proposed three general techniques for deep relevance training, including Siamese structure for pairwise relevance learning, efficient batch negative training, and model fine tuning through human labeled point-wise data. To study the effect of these techniques, we compare our model with several simpler versions.

The first set of experiments aim to evaluate the techniques proposed for training a model based on click data only. Particularly, we evaluate the following methods. 
\begin{itemize}
    \item{\textbf{Base Click Model (part)}} is the one trained on 10\% dataset, with both Siamese and batch negative techniques.
    \item{\textbf{w/o Siamese}} is a model without the Siamese structure, which takes a pairwise preference example ($Q$, $\rm{item}_a$, $\rm{item}_b$) as input, and outputs their relevance preference. As the model is trained to score a pairwise preference example, it's hard to evaluate point-wise relevance metrics(e.g., PR AUC) and we just report eval AUC and pairwise accuracy. 
     \item{\textbf{w/o Batch Negative}} is our base click model without batch negative examples. 
     \item{\textbf{DSSM}} is a classical deep matching model for web search~\cite{huang2013learning}. DSSM could be regarded as a simplified version of our model without the Siamese structure and batch negatives.
\end{itemize}
By comparing the results shown in Table-\ref{tbl:exp1}, the base click model boosts the performance on all metrics (outperforming DSSM by 13\% for pairwise accuracy, 28\% for precision-recall AUC, 13\% for negative PR-AUC and 3\% for NDCG), showing that our proposed general techniques facilitate the learning of relevance from click data and decrease the impact of distributional mismatches.

In the second set of experiments, we compare the proposed transfer learning technique with several variants and baselines.

\begin{itemize}
 \item{\textbf{GBDT}} (Gradient Boosted Decision Tree) is one of the most popular non-DL based models used in relevance learning ; see \cite{yin2016ranking} for detail. This is our in-house ranking model baseline, implemented with Xgboost \cite{xgboost}, with 60 features covering a variety of user feedback signals, hand-picked matching features, including BM25, window size, etc. This is optimized for years and is therefore a strong baseline.
\item{\textbf{Base Click Model (full)}} is trained on the whole dataset, with batch negative and Siamese structure, but without transfer learning. We take it as another baseline to highlight the effect of transfer learning. 
\item{\textbf{+Pointwise ensemble}} is our propose method.
\item{\textbf{+Pairwise ensemble}} is similar to our proposed approach but uses pairwise human label training target.
\item{\textbf{+Pointwise simple}} is similar to the proposed method but without the base click network.
\end{itemize}
As shown in Table \ref{tbl:exp2}, the transfer learning models outperform GBDT and the base click model in all the metrics. 

Ensembling improves the point-wise transfer learning model on all metrics. Pairwise ensemble on the other hand does the best on pairwise accuracy on the validation set, as expected. However, since PR-AUC and negative PR-AUC relate directly to the way we apply the model online, namely filtering of the top 100 results, transfer learning with the point-wise ensemble is our choice (outperforming GBDT by 3.6\% on negative PR-AUC).

%We compare the proposed model with our in-house state-of-the-art GBDT model and different transfer learning configurations. Gradient Boosted Decision Tree(GBDT) is one of the most popular non-DL based models used for learning relevance in search systems; see \cite{yin2016ranking} for detail. This is our in-house ranking model baseline, with 60 features covering a variety of user feedback signals, hand-picked matching features, including BM25, window size, etc. This is a strong baseline we would compare to. Base click model(full) is trained on the whole dataset, with batch negative and siamese structure. We take it as another baseline to compare with the one importing transfer learning.

\subsection{Case Studies}

To illustrate the power of batch negative co-training, we look at example queries and items and how they score under the models trained with and without batch negatives (vanilla model), both under the Siamese pairwise architecture.

We use the query ``cellphone" and the top 3 results returned by our commercial search engine, two of which are Huawei phones and the third one is an Apple iPhone X. We then change the query to ``paper cup", which is something completely unrelated and chosen randomly.

Under the session-pairs only model (first 2 columns in Table~\ref{case-studies}), the three smart phones do receive higher scores under ``cellphone" than under ``paper cup", but the score ranges are very similar. In contrast, the batch-negative enhanced model produces 10 times greater score separation under the two queries, without significantly dilating the cluster score range itself. Thus the latter model exhibits much greater confidence in scoring and penalizes comically unrelated (query, item) pairs, in accordance with intuition.